\documentclass[12pt]{article}
\begin{document}
\pagestyle{plain} \setcounter{page}{1}
\begin{center}
{\large \textbf{Variable Speed of Light Cosmology: An Alternative to
Inflation}}\vskip 0.2 true in
{\large J. W. Moffat}\footnote{e-mail: john.moffat@utoronto.ca}
\vskip 0.2 true in
\textit{Department of Physics, University of Toronto, Toronto, Ontario M5S
1A7, Canada} \vskip 0.2 true in
and
\vskip0.2 true in
\textit{Perimeter Institute for Theoretical Physics, Waterloo, Ontario N2J
2W9, Canada}
\vskip 0.2 true in
\begin{abstract} It is generally believed
that inflationary cosmology explains the isotropy, large scale homogeneity
and flatness as well as predicting the deviations from homogeneity of our
universe. We show that this is not the only cosmology which can explain
successfully these features of the universe. We consider anew and modify a
model in which local Lorentz invariance is spontaneously broken in the
very early universe, and in this epoch the speed of light undergoes a
first or second order phase transition to a value $\sim$ 30 orders of
magnitude smaller, corresponding to the presently measured speed of light.
Before the phase transition at a time $t\sim t_c$, the entropy of the
universe is reduced by many orders of magnitude, allowing for a
semiclassical quantum field theory calculation of a scale invariant
fluctuation spectrum. After the phase transition has occurred, the
radiation density and the entropy of the universe increase hugely and the
increase in entropy follows the arrow of time determined by the
spontaneously broken direction of the vev $\langle\phi^a\rangle_0$. This
solves the enigma of the arrow of time and the second law of
thermodynamics. A new calculation of the primordial Gaussian and adiabatic
fluctuation spectrum is carried out, leading to a scale invariant scalar
component of the power spectrum. We argue that there are several
attractive features of VSL theory compared to standard inflationary
theory, and that it provides an alternative cosmology with potentially
different predictions. \end{abstract}  \vskip 0.3 true in \end{center}

\date{\today}

\section{\bf Introduction}

It is ten years ago that an alternative solution to the initial value
problems of cosmology based on a variable speed of light (VSL) was
published~\cite{Moffat}. The model was based on the idea
that in the very early universe at a time $t\sim t_P\sim 10^{-43}$ sec.,
where $t_P$ denotes the Planck time, the local Lorentz invariance of the
ground state of the universe was spontaneously broken by means of a
non-zero vacuum expectation value (vev) of a vector field,
$\langle\phi^a\rangle_0 \not=0$, where $a$ labels the flat tangent space
coordinates of four-dimensional spacetime. At a temperature $T < T_c$, the
local Lorentz symmetry of the vacuum was restored corresponding to an
``anti-restoration'' of the symmetry group $SO(3,1)$.  Above $T_c$ the
symmetry of the ground state of the universe was broken from $SO(3,1)$
down to $O(3)$, and the domain formed by the direction of
$\langle\phi^a\rangle_0$ produced an arrow of time pointing in the
direction of increasing entropy and the expansion of the universe.

The notion that as the temperature of the universe increases, a larger
symmetry group $SO(3,1)$ can spontaneously break to a smaller group $O(3)$
seems counter-intuitive. Heating a superconductor restores gauge
invariance, and heating a ferromagnet restores rotational invariance.
Anti-restoration would appear to violate the second law of thermodynamics.
This, however, is not the case, for certain ferroelectric crystals such as
Rochelle or Seignette salt, possess a smaller invariance group above a
critical temperature, $T=T_c$, than below it~\cite{Jona}\footnote{Rochelle
salt possesses a lower Curie point at $-18^o\,C$, below which the
Rochelle crystal is orthorhombic and above which it is monoclinic.}
Explicit models of 4-D field theories have been constructed in which the
symmetry non-restoration of symmetries occurs at high
temperatures~\cite{Weinberg}\footnote{Temperature dependent models of
spontaneous symmetry breaking of $SO(3,1)\rightarrow O(3)$ with increasing
temperature will be investigated in a separate article.}

Recently, Hollands and Wald~\cite{Wald} investigated the issue of whether
an alternative to inflation theory could exist and predict a
scale invariant fluctuation spectrum in the early universe. They also
considered the issue of fine-tuning in inflationary cosmology, and argued
that the present models of inflation do not avoid fine-tuning of the
initial conditions of the universe, although they do improve considerably
the extreme fine-tuning that occurs in the standard Friedmann,
Robertson and Walker (FRW) big bang model. They stressed that an important
issue in the initial value problem in cosmology is to explain how
the second law of thermodynamics came into being.

Kofman, Linde
and Mukhanov~\cite{Mukhanov} argued that inflation theory is not as
fine-tuned in the initial universe as claimed by Hollands and Wald,
although the avoidance of such a fine-tuning relies to some extent on the
use of the ``anthropic principle'', in that chaotic inflation postulates
enough initial patches of potential inflation such that one of them can
develop enough e-folds of inflation to solve the horizon and flatness
problems in our universe~\cite{Linde}. They also stressed that without a
scenario such as inflation in the early universe, it would not be possible
to dilute the initial density of radiation and matter to allow a sensible
semi-classical quantum field theory calculation of quantum fluctuations.
Indeed, the fluctuation calculation of Hollands and Wald would have to be
performed at a density, $\rho\sim 10^{95}\,\rho_P$, where
$\rho_P=c^5/\hbar G^2$ is the Planck density, a density so large that it
would not allow any standard quantum field theory calculations to be
carried out. Inflation models do significantly dilute the radiation and
matter density of the early universe and they also exponentially reduce
the entropy of the universe, leading to a resolution of the flatness
problem. However, the matter density and entropy have to be re-instated by
a period of re-heating in which the inflaton field and the large vacuum
energy undergo decay.

This then leads us inevitably to the question: Does there exist an
alternative to inflation, which can successfully allow a quantum field
theory calculation of a scale invariant primordial spectrum? In spite of
the successes of inflation theory, it is important to seek alternatives to
it to see whether a different scenario could overcome some of the
shortcomings of inflation, such as the problem of vacuum energy, the
fine-tuning of the coupling constant to give the correct density profile in
the present universe, and the unnaturally flat potentials needed to solve
the initial value problems. In the following, we shall consider anew the
VSL cosmology associated with a spontaneous
symmetry breaking of Lorentz invariance and a phase transition in the
speed of light in the very early universe.

Alternative VSL models, such as
those considered by Albrecht and Maguiejo, and
Barrow~\cite{Magueijo} were based on a ``hard'' breaking of Lorentz
invariance. We, instead, attempt to emulate the successes of the standard
model of particle physics~\cite{Cheng}, in which ``soft'', spontaneous
breaking of the internal symmetries by a Higgs mechanism plays a crucial
role. In our scenario, local Lorentz symmetry is simply an accident of
nature, i.e. the ground state of our current universe happens to be found
in a particular false vacuum, and transitions away from this ground state
may well have happened in the early universe.

Another alternative model of VSL theory has been based on a bimetric theory
of gravity~\cite{Clayton,Clayton2,Moffat2}. In this model there are two
metrics in which the light cones are associated, respectively, with the
speed of light and the speed of gravitational waves, and they are linked
by the gradient of a scalar field. These models maintain local
relativistic invariance and diffeomorphism invariance, and do resolve the
early universe initial value problems. The basic parameter in these models
is the {\it dimensionless} ratio of the speed of gravitational waves
(graviton) to the speed of light (photon). However, a calculation of the
primordial scale invariant fluctuation spectrum has, so far, only been
performed by using a ``slow roll'' approximation for the potential of the
scalar field that links both metrics~\cite{Clayton2}. Such a scheme falls
into the category of inflationary cosmology. Magueijo~\cite{Magueijo2} has
also published a version of a relativistic VSL model, but this model has
not yet succeeded in producing a viable scale invariant fluctuation
spectrum, which must play a crucial role in confirming models of the early
universe.

A vierbein ${e_\mu}^a$ is used to convert $\phi^a$ into a 4-vector in
coordinate space: $\phi^\mu={e_a}^\mu\phi^a$ and it satisfies\footnote{The
indices $\mu,\nu...$ and $a,b...$ run from $0,...,3$ and the Minkowski
metric signature is $\eta_{\mu\nu}={\rm diag}(1,-1,-1,-1).$}
\begin{equation}
e^a_\mu e^\mu_b=\delta^a_b,\quad e^\mu_ae^a_\nu=\delta^\mu_\nu.
\end{equation}
The vierbein obeys the Lorentz transformation rule
\begin{equation}
e'^a_\mu(x)=L^a_b(x)e^b_\mu(x).
\end{equation}
The metric tensor is obtained from the vierbeins by the formula
\begin{equation}
g_{\mu\nu}=\eta_{ab}e^a_\mu e^b_\nu.
\end{equation}

The covariant derivative operator acting on $\phi^a$ is defined by
\begin{equation}
D_\mu\phi^a=[\partial_\mu\delta^a_b+(\Omega_\mu)^a_b]\phi^b,
\end{equation}
where $(\Omega_\mu)^a_b$ denotes the spin, gauge connection:
\begin{equation}
\Omega_\mu=\frac{1}{2}\sigma^{ab}e^\nu_a\nabla_\mu e_{b\nu},
\end{equation}
and $\nabla_\mu$ denotes covariant differentiation with respect to the
Christoffel symbol $\Gamma^\lambda_{\mu\nu}$:
\begin{equation}
\Gamma^\lambda_{\mu\nu}=g^{\lambda\rho}\eta_{ab}(D_\mu e^a_\nu)e^b_\rho.
\end{equation}
Moreover, the $\sigma_{ab}$ are the six generators of the Lorentz group.

We shall describe the dynamical behavior of the speed of light $c(x)$ by
a scalar field: $c(x)={\bar c}\chi(x)$ where ${\bar c}$ is a constant
with the dimensions of velocity. The total action of the theory is
\begin{equation}
S=S_G+S_M+S_\phi+S_\chi,
\end{equation}
where
\begin{equation}
\label{grav}
S_G=-\frac{c^4}{16\pi G}\int d^4xe(R+2\Lambda),
\end{equation}
and $e=\sqrt{-g}={\rm det}(e^a_\mu e_{a\nu})$, $\Lambda$ is the
cosmological constant and $S_M$ is the matter action. Moreover,
\begin{equation}
S_\phi=\int
d^4x\sqrt{-g}\biggl[\frac{1}{2}D_\mu\phi_aD^\mu\phi^a-V(\phi)\biggr].
\end{equation} We demand that $\phi^a$ (or $\phi^\mu$) be a timelike
vector, which ensures that the kinetic energy term $D_\mu\phi_aD^\mu\phi^a
>0$ for all events in the past and future light cones of the flat tangent
space, which avoids the occurrence of negative energy modes in the
Hamiltonian. We could add a Lagrange multiplier term to the action to
guarantee the timelike nature of the vector
$\phi^a$~\cite{Dirac}\footnote{An alternative proposal would replace the
kinetic energy term $\frac{1}{2}D_\mu\phi_aD^\mu\phi^a$ by
$\frac{1}{2}(D_\mu\phi^\mu)^2$, which would avoid any negative energy ghost
states~\cite{Clayton3}. This proposal will be investigated in a future
publication.} We also have
\begin{equation}
S_\chi=\int d^4x\sqrt{-g}\biggl[\frac{1}{2}D_\mu\chi
D^\mu\chi-V(\chi)-V(\chi\phi)\biggr],
\end{equation}
where $V(\phi\chi)$ denotes a potential energy contribution coupling the
fields $\phi^\mu$ and $\chi$ (e.g. a Yukawa coupling contribution $\partial_\mu\phi^\mu\chi$).

We choose the potential $V(\phi)$ to be
of the form
\begin{equation}
V(\phi)=-\frac{1}{2}\mu^2\phi_a\phi^a+\lambda(\phi_a\phi^a)^2,
\end{equation}
where $\phi_a\phi^a > 0$ and the coupling constant $\lambda
> 0$, so that the potential is bounded from below. If $V$ has a minimum at
$\phi_a=v_a$, then the spontaneously broken solution is given by
$v_a^2=\mu/4\lambda$. We can choose $\phi_a$ to be
\begin{equation}
\phi_a=\delta_{a0}v=\delta_{a0}(\mu^2/4\lambda)^{1/2}.
\end{equation}
All the other solutions of $\phi_a$ are related to this one by a Lorentz
transformation. Then, the homogeneous Lorentz group $SO(3,1)$ is broken
down to the spatial rotation group $O(3)$. The three rotation generators
$J_i\,(i=1,2,3)$ leave the vacuum invariant, $J_iv_i=0$, while the three
Lorentz-boost generators $K_i$ break the vacuum symmetry, $K_iv_i\not= 0$.

Let us consider small oscillations about the true minimum and define a
shifted field $\phi'_a=\phi_a-v_a.$ By performing a Lorentz transformation,
we obtain
\begin{equation}
\phi^0=\psi,\quad \phi^1=\phi^2=\phi^3=0.
\end{equation}
In this special coordinate frame, the remaining component $\psi$ is the
scalar physical particle that survives after the three Goldstone modes have
been removed. This corresponds to the ``unitary gauge'' in the standard
electroweak theory. In the broken phase, the Einstein tensor
$G^{\mu\nu}=R^{\mu\nu}-\frac{1}{2}g^{\mu\nu}R$ still satisfies the Bianchi
identities: $\nabla_\nu G^{\mu\nu}\equiv 0$, but the conservation law for
the energy momentum tensor is modified to be~\cite{Moffat}:
\begin{equation}
\nabla_\nu T^{\mu\nu}=-\nabla_\nu(K^{\mu\nu}+H^{\mu\nu}),
\end{equation}
where $K^{\mu\nu}$ and $H^{\mu\nu}$ are non-vanishing contributions that
arise in the spontaneously broken phase due to the ``Higgs mechanism'' for
the spin gauge field $\Omega_\mu$, and the energy momentum tensor for the
physical fields $\psi$ and $\chi$, respectively. In the unbroken phase, we
regain the standard energy momentum conservation law $\nabla_\nu
(T^{\mu\nu}+H^{\mu\nu})=0$, since $K^{\mu\nu}=0$ and the
spin connection becomes that of a massless graviton gauge field.

\section{\bf Variable Speed of Light and Solutions to the Horizon and Flatness Problems}

In the spontaneously broken phase of the evolution of the universe, the
spacetime manifold has been broken down to $R\times O(3)$. The
three-dimensional space with $O(3)$ symmetry is assumed to be the
homogeneous and isotropic FRW solution:
\begin{equation}
d\sigma^2=R^2(t)\biggl[\frac{dr^2}{1-kr^2}+r^2(d\theta^2+\sin^2\theta
d\phi^2)\biggr],
\end{equation}
where $k=0,+1,-1$ corresponding to a flat, closed and open universe,
respectively, and $t$ is the external time variable. This describes the
space of our ordered ground state in the symmetry broken phase and it has
the correct subspace structure for our FRW universe with the metric
\begin{equation}
\label{metric}
ds^2\equiv
g_{\mu\nu}dx^\mu
dx^\nu=dt^2c^2(t)-R^2(t)\biggl[\frac{dr^2}{1-kr^2}+r^2(d\theta^2+\sin^2\theta
d\phi^2)\biggr]. \end{equation} The Newtonian ``time'' $t$ is the {\it
absolute time} measured by standard clocks.

In the spontaneously broken Lorentz symmetry phase, we can now have the
speed of light $c$ undergo a phase transition, since we are no longer
required to satisfy Einstein's second postulate of special relativity: The
speed of light $c$ is the same constant with respect to all observers
irrespective of their motion and the motion of the source.

Close to the phase transition at the time $t\sim t_c$, we assume that
\begin{equation}
c(t)=c_0\theta(t_c-t)+c_m\theta(t-t_c),
\end{equation}
where $\theta(t)$ is the Heaviside step function which satisfies
$\theta(t)=1$ for $t>0$ and $\theta =0$ for $t <0$. Moreover, $c_0$ and
$c_m$ denote the values of $c$ before and after the phase transition,
respectively, where $c_m=299792458\,m\,s^{-1}$ is the presently
measured value of $c$ and $c_0\gg c_m$.

The metric $g_{\mu\nu}$ now has the bimetric form:
\begin{equation}
g_{\mu\nu}=g_{0\,\mu\nu}+g_{m\,\mu\nu},
\end{equation}
where
\begin{equation}
\label{brokenmetric}
ds_0^2\equiv
g_{0\,\mu\nu}dx^\mu
dx^\nu=dt^2c_0^2\theta(t_c-t)-R^2\biggl[\frac{dr^2}{1-kr^2}+r^2(d\theta^2+\sin^2\theta
d\phi^2)\biggr],
\end{equation}
and
\begin{equation}
\label{spmetric}
ds^2_m\equiv
g_{m\,\mu\nu}dx^\mu
dx^\nu=dt^2c_m^2\theta(t-t_c)-R^2\biggl[\frac{dr^2}{1-kr^2}+r^2(d\theta^2+\sin^2\theta
d\phi^2)\biggr].  \end{equation}

The phase transition in $c$ produces two light cones: $ds_0^2=0$ and
$ds_m^2=0$ and their relative sizes are determined by the {\it
dimensionless} ratio $\gamma=c_0/c_m$. When
$\gamma=1$ and there is no phase transition the model becomes the same as
local special relativity with one light cone.  When $\gamma$ becomes very
large in the spontaneously broken phase, the Minkowski light cone,
determined by the metric (\ref{spmetric}), is contained within the much
larger light cone determined by the metric (\ref{brokenmetric}). As in
alternative bimetric theories, a diffeomorphism transformation in time
cannot remove the speed of light dependence from both metrics
$g_{0\,\mu\nu}$ and $g_{m\,\mu\nu}$ simultaneously. Only when $\gamma=1$
can a diffeomorphism time transformation $dt'=dtc(t)$ remove $c(t)$
completely, for then we have only one light cone and one speed of light
corresponding to local special relativity.

The proper horizon scale is
given by
\begin{equation}
d_H(t)=R(t)\int_0^t\frac{dt'c(t')}{R(t')}.
\end{equation}
We obtain for $t>t_c$ the usual result, $d_H\sim 2c_mt$, since for a
radiation dominated universe $R(t)\propto t^{1/2}$ and $c=c_m$. On the
other hand, we have for a radiation dominated universe in the
spontaneously broken phase before the phase transition in $c(t)$,
$d_H\sim 2c_0t$, and for $\gamma\rightarrow \infty$
the proper horizon size is stretched and this means that all observers in
the spontaneously broken phase were in causal contact. The forward light
cone beginning at the time of the big bang is considerably expanded for
$\gamma\rightarrow\infty$ and is made larger than the region from which
microwave photons are reaching us today and this solves the isotropy
problem.

To see how the flatness problem is resolved, we write the Friedmann
equation in the spontaneously broken phase:
\begin{equation}
H^2+\frac{c^2k}{R^2}=\frac{8\pi G\rho}{3}+\frac{c^2\Lambda}{3},
\end{equation}
where $H={\dot R}/R$. We set the cosmological constant $\Lambda=0$, and
obtain
\begin{equation}
\epsilon\equiv\vert\Omega-1\vert=\frac{c^2\vert k\vert}{{\dot R}^2},
\end{equation}
where $\Omega=8\pi G\rho/3H^2$. We now find that
\begin{equation}
{\dot\epsilon}=-\frac{2c^2\vert k\vert{\ddot R}}{{\dot R}^3}
+2\biggl(\frac{\dot c}{c}\biggr)\biggl(\frac{c^2\vert k\vert}{{\dot
R}^2}\biggr).
\end{equation}
For a radiation dominated universe ${\ddot
R}<0$ and for a speed of light $c$ that decreases in a phase transition
to the small value $c_m$, we have ${\dot c}/c< 0$ and ${\dot\epsilon} <
0$ corresponding to an attractor solution with $\epsilon\sim 0$ and an
approximately spatially flat universe.

An alternative way of seeing how the
flatness problem is resolved is to write
\begin{equation}
\Omega(t)=1+x(t),
\end{equation}
where
\begin{equation}
x(t)=\frac{c^2k}{R^2H^2}\sim\frac{c^2k/R^2}{8\pi G\rho_r/3},
\end{equation}
where $\rho_r$ is the radiation density,
$\rho_r=\rho_{0r}\biggl(R_0/R\biggr)^4$ with $\rho_{0r}$ and $R_0$ denoting
the present values of the radiation density and $R$, respectively. Then,
we have $x\sim c^2kR^2/R^*$ where $R^*=8\pi G\rho_{0r}R_0^4/3$. This
yields close to the phase transition with $\gamma=c_0/c_m$:
\begin{equation}
\vert\Omega(10^{-43}\,{\rm sec})-1\vert\sim {\cal O}(\gamma^210^{-60}).
\end{equation}
Thus, in the time the universe is in the broken phase before the phase
transition in the speed of light, we obtain for $\gamma\sim 5\times
10^{29}$:
\begin{equation}
\vert\Omega(10^{-43}\,{\rm sec})-1\vert\sim{\cal O}(1),
\end{equation}
which implies much less fine-tuning than the standard FRW model.

\section{\bf The Entropy Problem and the Arrow of Time}

A calculation of the energy density of photons gives
\begin{equation}
{\cal E}_\gamma=\sigma_BT^4,
\end{equation}
where $\sigma_B$ is the
Stefan-Boltzmann constant, $\sigma_B=\pi^2k_B^4/15c_m^3\hbar^3=7.5641\times
10^{-15}\,{\rm erg}\,{\rm cm}^{-3}\,K^{-4}$ ($k_B$ is Boltzmann's
constant). For $\log_{10}\gamma \sim 30$, we see that
the Stefan-Boltzmann constant is significantly reduced to:
$\sigma_B=\pi^2k^4_B/15c_0^3\hbar^3\sim 7.56\times 10^{-104}\,{\rm
erg}\,{\rm cm}^{-3}\,K^{-4}$. Thus, in the early universe in the
symmetry restored phase for a temperature $T\sim 10^{12}\,K$, we have
${\cal E}_\gamma\sim 7.6\times 10^{33}\,{\rm erg}\,{\rm cm}^{-3}$ whereas
in the spontaneously broken phase the energy density of photons is
significantly diluted,  ${\cal E}_\gamma\sim 400\,{\rm erg}\,{\rm
cm}^{-3}$.

The entropy of relativistic particles is given by~\cite{Weinberg2}:
\begin{equation}
S=\frac{R^3}{T}(\rho c^2+p)=\frac{4\sigma_B}{3}(RT)^3f,
\end{equation}
where $f$ is a numerical factor of order unity. For $\gamma=1$, $T\sim
10^{12}\,K$ and $R\sim 10^{28}\, {\rm cm}$ we get $S\sim 10^{118}\,{\rm
erg}\,K^{-1}$. In the broken symmetry phase, near the phase transition in
$c$, and for $\log_{10}\gamma\sim 30$ this will be reduced to the small
entropy, $\sim 300\,{\rm erg}\,K^{-1}$. The phase transition in the speed
of light with $\gamma\rightarrow 1$, results in an enormous increase in the
entropy of the universe.

A similar stuation occurs in inflationary models in which the exponential
expansion of the universe decreases considerably the entropy and results in
a re-heating phase when inflation ceases.

In our VSL scenario, the large increase in entropy at the phase transition
time $t\sim t_c$ is in the direction of the spontaneous symmetry breaking
domain $\langle\phi^a\rangle_0\not= 0$, which corresponds to the direction
of the arrow of time in the expanding universe. To solve the problem of the
arrow of time and the second law of thermodynamics, we should expect that
the entropy of the universe at or near the big bang should be
small~\cite{Penrose}. The sudden large increase in entropy in our VSL
scenario also leads to a solution of the flatness problem, as in the case
of inflationary models.

In the spontaneously broken phase near the phase transition
$c_0\rightarrow \infty$, the Planck length $L_P=\sqrt{\hbar
G/c_0^3}\rightarrow 0$, and the Planck density $\rho_P=c_0^5/\hbar
G^2\rightarrow \infty$. Thus, the super-Planck density is far
removed from the region in the spontaneously broken phase, in which the
radiation energy ${\cal E}_\gamma$ and the entropy $S$ are diluted as
$c_0\rightarrow\infty$, and we do not have to concern ourselves with
ultra-Planck energy corrections to the primordial fluctuation spectrum,
which is calculated in the spontaneously broken phase.

\section{\bf Calculation of Scale Invariant Fluctuation Spectrum}

Inflationary models provide a simple and successful answer to how the
departures from inhomogeneity arise from quantum fluctuations in the early
universe. This prediction of a scale invariant spectrum has been
confirmed during the past two years by high precision
measurements of the cosmic microwave background (CMB)~\cite{Netterfield}.
We shall now show how our VSL model can predict equally well a scalar,
adiabatic Gaussian scale-free perturbation spectrum. We shall use a simple
method for calculating the spectrum, avoiding many of the technical
details, so that we can see how the mechanism works at an intuitive
level~\cite{Wald}.

We shall consider a simple model of a free, minimally coupled
scalar field $\psi$, which we identify with our physical field $\psi$ in
the ``unitary gauge'' after the three Goldstone modes have been removed in
our model of spontaneous symmetry breaking of Lorentz invariance. We choose
for simplicity the flat spacetime with $k=0$.
The scalar field $\psi$ is pictured as a
plane wave mode with coordinate wave vector ${\vec k}$:
\begin{equation}
\psi({\vec x},t)=\psi_k(t)\exp(i{\vec k}\cdot{\vec x}),
\end{equation}
which satisfies
\begin{equation}
\label{Harmonic}
{\ddot\psi}_k+3H{\dot\psi}_k+\frac{c^2k^2}{R^2}\psi_k=0,
\end{equation}
and we have defined
\begin{equation}
\psi_k=\frac{1}{(2\pi)^{3/2}}\int d^3x\psi({\vec x})\exp(-i{\vec
k}\cdot{\vec x}).
\end{equation}
Here, we consider that the quantum fluctuation modes are created in the
spontaneously broken ground state and that their proper wavelength
$\lambda_p$ is tiny compared to the Hubble radius, $R_H=c/H$. The
equation of motion for the dynamical field $\chi(t)$ in the preferred frame gauge,
$\phi^0=\psi$, is of the form:
\begin{equation}
\label{cequation}
{\ddot\chi}+3H{\dot\chi}+\frac{dV(\chi)}{d\chi}+\frac{dV(\chi\psi)}{d\chi}+I(\chi,g)=0,
\end{equation}
where $I(\chi,g)$ denotes the contribution coming from the variation of
$\chi$ in the Einstein-Hilbert action $S_G$ in (\ref{grav}).  A possible
solution for $c(t)={\bar c}\chi(t)$ given the potentials $V(\chi)$,
$V(\phi\chi)$ and $I(\chi,g)$ is
\begin{equation}
\label{csolution}
c(t)=\frac{a}{t^b}+c_0\theta(t_c-t)+c_m\theta(t-t_c),
\end{equation}
where $c(t)\rightarrow c_0$ from above as $t\rightarrow t_c$.

Eq.(\ref{Harmonic}) has the same form as the harmonic oscillator equation
with a unit mass, a variable spring constant $c^2k^2/R^2$, and a variable
friction damping coefficient $3H$. The Lagrangian for our harmonic oscillator has the form
\begin{equation}
L_k=\frac{R^3}{2}\biggl({\dot\psi}_k^2-\frac{c^2k^2}{R^2}\psi_k^2\biggr).
\end{equation}
The ground state of the oscillator at some fixed time $t$ has the form
of a Gaussian wave function, with a spread given by
\begin{equation}
\label{spread}
(\Delta\psi_k)^2=\frac{1}{2R^2ck}.
\end{equation}

In the case of generic inflationary models, we have $R(t)\sim \exp(Ht)$ and
$H$ is constant in time. When the proper wavelength,
$\lambda_p=R/k$, of the normal mode is much smaller than the Hubble radius
$R_H=c_m/H$, the mode oscillates like an ordinary harmonic oscillator with
small damping, and the adiabatic vacuum corresponds to that of a
flat Minkowski spacetime. However, when $\lambda_p$ is much greater than
$R_H$ the mode enters an overdamped phase with ${\dot\psi}_k\sim 0$ and
the mode ``freezes''. Indeed, we have that $\lambda_p=R/k\sim \exp(Ht)$,
while the Hubble radius $R_H$ remains constant during the inflationary
period, so that the proper wavelengths of the normal modes quickly
overtake the horizon and make a frozen imprint on the spacetime metric. On
the other hand, in the standard FRW model for the radiation equation of
state, $p=\frac{1}{3}\rho$, we obtain $\lambda_p\propto t^{1/2}$ while
$R_H\sim 2c_mt$, so that the proper wavelengths of the initial tiny wave
modes {\it never catch up to the Hubble horizon and cross it} to produce a
scale-free fluctuation spectrum.

For our spontaneously broken VSL model, we have from (\ref{Harmonic})
and (\ref{csolution}) for times $t < t_c$ and for a radiation dominated
background universe with $R\sim At^{1/2}$ and $H\sim 1/2t$:
\begin{equation}
\label{psiequation}
{\ddot\psi}_k+\frac{3}{4t^2}\psi_k+\biggl(\frac{a}{A}\biggr)^2\frac{k^2}{t^{2b+1}}\psi_k=0,
\end{equation} where we have chosen ${\dot\psi}_k\sim H\psi_k\sim
(1/2t)\psi_k$. We observe that as $t\rightarrow 0$ for $b \geq 1$, there
will be a period in the spontaneously broken phase in which
(\ref{psiequation}) produces oscillating modes, for the wavelengths
$\lambda_p\sim t^{1/2}$ are much smaller than the Hubble radius, $R_H\sim
c/H\sim a/t^{b-1}$. At this time, the universe evolves adiabatically {\it
in a Minkowski flat spacetime vacuum} and the ground state remains more or
less as in (\ref{spread}). However, as the universe expands and $t$
increases, the wavelengths $\lambda_p\sim t^{1/2}$ will overtake the
Hubble radius $R_H\sim a/t^{b-1}$ and cross it as $t$ approaches the
phase transition time $t_c$. The overdamped modes cease to oscillate and
$\Delta\psi_k$ will become constant at the fixed time $t=t_h$ when the
wavelengths cross the horizon.

After the comoving wavelengths pass through the horizon, they freeze and
the spectrum spread is given by
\begin{equation}
\label{freeze}
(\Delta\psi_k)_h^2=\frac{1}{2R_h^2c_hk},
\end{equation}
where $c_h$ and $R_h$ are the values of $c$ and $R$ at the time the
modes cross the Hubble radius, $R_h=c_h/H_h$, i.e. when
\begin{equation}
\label{horizon}
\frac{R_h}{k}=\frac{c_h}{H_h}.
\end{equation}
Therefore, the fluctuation modes at later times have the spectrum
\begin{equation}
(\Delta\psi_k)_h^2\sim \frac{H_h^2}{c_h^3k^3}.
\end{equation}
The horizon radius $R_h$ can be made to coincide with the phase transition
with $R_h\sim c_{\rm ph}/H_{\rm ph}$ when $H_{\rm ph}$ is expected to
be approximately constant.

This constitutes the prediction of a scale
invariant spectrum with
\begin{equation}
k^3\vert\delta_k\vert\sim {\rm constant},
\end{equation}
where $\delta_k$ is the fractional energy density fluctuation in momentum
space. We observe that the difference between (\ref{spread}) and (\ref{freeze})
at later times is given by
\begin{equation}
\biggl(\frac{R}{R_h}\biggr)^2\biggl(\frac{c_0}{c_h}\biggr)(\Delta\psi_k)^2
=(\Delta\psi_k)_h^2.
\end{equation}
We see that for $c_h=c_m$ the spread $(\Delta\psi_k)^2$ is magnified by the
huge factor $\log_{10}\gamma\sim 30$, so that the late time quantum
fluctuations have macroscopically relevant cosmological interest. In
inflation theory, it is the factor $(R/R_h)^2$ that is exponentially
enhanced and also produces macroscopically large fluctuation effects.

\section{\bf Comparison of VSL and Inflationary Cosmologies}

Let us compare the VSL and inflationary models. Regarding the problem of
fine-tuning of the initial conditions after the big bang, the argument
given by Hollands and Wald~\cite{Wald} that the initial conditions in
inflation cannot be natural, depends on whether inflation models can be
considered time reversible, so that the probability that a universe would
get large by undergoing an era of inflation is equal to the probability
that a universe will undergo an era of ``deflation'' when it recollapses.
It is argued that the probability that a universe dominated by ordinary
matter will deflate is very small, so that by time invariance the
probability of inflation must be small too.

It is argued by Kofman, Linde and Mukhanov~\cite{Mukhanov} that the
dynamical evolution of the universe does not preserve the measure of
probability, or the number of degrees of freedom. This is mainly due to the
circumstance that in inflationary models the total energy of the scalar
inflaton field and the particles created by its decay is not conserved.
Since all the $\sim 10^{88}$ particles we see now within our horizon were
created by the scalar inflaton, then inflation had removed them at the
beginning of the universe and thereby guaranteed the absence of
adiabaticity. Thus, inflationary evolution can never produce the same
initial conditions at the universe's beginning. This circumstance should be
considered in contrast to the fact that the equations of Einstein's general
relativity are time reversible invariant, so that there is an equal number
of decreasing and growing entropy universes.

What can we say about the likelihood of a spontaneous symmetry breaking of
Lorentz invariance of the ground state occurring in the early universe, and
a sudden phase transition happening in the speed of light? The Lorentz
symmetry of the ground state of the universe is just an accident, for the
symmetry occurs in a false vacuum state and could occur at any time. It
is, of course, difficult to measure the probability of such an event, as
it would be for the spontaneous breaking of the internal symmetries of the
standard model.

A fundamental difference between VSL cosmology and inflation is that we
can choose the cosmological constant $\Lambda$ to be small or zero from the
beginning of the universe. If the data supporting an accelerating expansion
of the universe continues to be affirmed by more
observations~\cite{Perlmutter}, then we can have a small positive
cosmological constant in the present universe. In inflationary models, the
initial vacuum energy coming from the inflaton potential is huge, so that
enough e-folds of inflation can be sustained. How do we know that the
long-sought mechanism for the explanation of the smallness of the
effective cosmological constant will not cancel out the large vacuum energy
needed in the inflationary era?~\cite{Brandenberger}. Moreover, the
decaying vacuum energy has to be fine-tuned to fit the present
observational data supporting a small positive cosmological constant.

In our VSL model, the vacuum energy does not play a crucial role in
solving the initial value problems. The pressure in our perfect
fluid model can always be positive, i.e. for the equation of state for
radiation and matter, $p=w\rho$, we can have $0\leq w\leq 1/3$ in the early
universe. Therefore, for dark matter and radiation there is no violation
of the positive energy conditions. However, if there is dark energy
causing an acceleration of the present universe, then the equation of state
for dark energy would be $p_{\rm DE}=w_{\rm DE}\rho_{\rm DE}$ and $-1 \leq
w_{\rm DE}<-2/3$.

Another feature associated with generic inflationary models is the extreme
flatness of the inflaton potential required to permit sufficient inflation
to occur. Apart from the necessary ``Mexican hat'' potential required to
allow for the spontaneous symmetry breaking of local Lorentz
invariance of the vacuum, the VSL model does not require a fine-tuning of
potentials of the kind needed by inflationary models. Of course, we do
not yet possess a microscopic model of the phase transition in the speed
of light, but we do possess an effective theory that can be modelled in
analogy with a semi-classical description of phase transitions in critical
phenomena.

\section{\bf Conclusions}

We have investigated anew and modified a model of VSL cosmology, first
published a decade ago, and compared it with standard inflationary
cosmology. A new calculation of the scale invariant fluctuation spectrum
agrees well with the data~\cite{Bond}, in the same way as the equivalent
calculation of the spectrum in inflation models. However, a more technical
calculation of the spectrum, including a derivation of the tensor component
needs to be performed~\cite{Brandenberger2}. It is expected that the tensor
and gravitational wave component of the fluctuation spectrum will not
necessarily agree with that predicted by inflation, providing a new
competitive prediction to be tested by observations.

We cannot, of course, go back to the beginning of the
universe to observe whether it actually went through an era of
exponential or power law inflation, or an epoch in which the
Lorentz invariance of the ground state of the universe was spontaneously
broken, accompanied by a phase transition in the speed of light.
Therefore, we must rely on the self-consistent results of calculations of
the primordial power spectrum and observations of CMB anisotropies to guide
us in our understanding of early universe cosmology.

We believe that the VSL cosmology considered here is a viable alternative
to standard inflationary models, and that it may overcome certain
shortcomings in the latter models, and produce new predictions that could
be tested and compared with inflationary scenarios.

\vskip0.2 true in \textbf{Acknowledgments} \vskip0.2 true in
This work was
supported by the Natural Sciences and Engineering Research Council of
Canada. I thank Michael Clayton for helpful discussions.
\vskip0.5 true in


\begin{thebibliography}{99}

\bibitem{Moffat} J. W. Moffat, Int. J. Mod, Phys. {\bf D2}, 351 (1993),
gr-qc/9211020; J. W. Moffat, Found. of Phys. {\bf 23}, 411 (1993),
gr-qc/9209001; J. W. Moffat, {\it Fluctuating Paths and Fields},
Festschrift dedicated to Hagen Kleinert, edited by H. Jancke, A. Pelster,
H.-J. Schmidt, and M. Bachmann, World Scientific, Singapore, p. 741, 2001,
astro-ph/9811390.

\bibitem{Jona} F. Jona and G. Shirane, {\it Ferroelectric Crystals},
Pergamon, Oxford, 1962.

\bibitem{Weinberg} S. Weinberg, Phys. Rev. {\bf D19}, 3357 (1974); R. N.
Mohapatra and G. Senjanovic, Phys. Rev. Lett. {\bf 42}, 1651 (1979); Phys.
Rev. {\bf D20}, 3390 (1979); P. Langacker and S. Pi, Phys. Rev. Lett. {\bf
45}, 1 (1980); P. Salomonson and B. K. Skagerstam, Phys. Lett. {\bf B155},
100 (1985); S. Dodelson and L. M. Widrow, Phys. Rev. {\bf D42}, 326 (1990).

\bibitem{Wald} S. Hollands and R. M. Wald, gr-qc/0205058.

\bibitem{Mukhanov} L. Kofman, A. Linde and V. Mukhanov, hep-th/0206088.

\bibitem{Linde} A. H. Guth, Phys. Rev. {\bf D23}, 347 (1981); A. Albrecht
and P. Steinhardt, Phys. Rev. Lett. {\bf 48}, 1220 (1982); A. D. Linde,
Phys. Lett. {\bf B129}, 177 (1983); A. D. Linde, {\it Particle Physics and
Inflationary Cosmology}, (Harwood, Chur, Switzerland, 1990).

\bibitem{Magueijo} A. Albrecht and J.
Magueijo, Phys. Rev. {\bf D59}, 043516, (1999), astro-ph/9811018; J. D. Barrow, Phys.
Rev. {\bf D59}, 043515 (1999); J. D. Barrow and J. Magueijo, Phys.
Lett. {\bf B477}, 246 (1999), astro-ph/9811073 v2.

\bibitem{Cheng} See e.g.: Ta-Pei Cheng and Ling-Fong Li, {\it Gauge Theory
of Elementary Particle Physics}, Clarendon Press, Oxford, 1984.

\bibitem{Clayton} M. A. Clayton and J. W. Moffat, Phys.
Lett. {\bf B460}, 263 (1999), astro-ph/9812481; M. A. Clayton and J. W.
Moffat, Phys. Lett. {\bf B477}, 269 (2000), gr-qc/9910112; M. A. Clayton
and  J. W. Moffat, Phys. Lett. {\bf B506}, 177 (2001), gr-qc/0101126 v2; M.
A. Clayton and J. W. Moffat, Int. J. of Mod. Phys. {\bf D11}, 187 (2002),
gr-qc/0003070; ; I. T. Drummond, Phys. Rev. {\bf D63} 043503 (2001), astro-ph/0008234; B. A.
Bassett, S. Liberati, C. Molinari-Paris, and M. Visser, Phys. Rev. {\bf
D62} 103518 (2000), astro-ph/0001441 v2.

\bibitem{Clayton2} M. A. Clayton and J. W. Moffat, astro-ph/0203164.

\bibitem{Moffat2} J. W. Moffat, to be published in Int. J. Mod. Phys.
(2002), gr-qc/0202012.

\bibitem{Magueijo2} J. Magueijo, Phys. Rev. {\bf D62},
103521 (2000), gr-qc/0007036.

\bibitem{Dirac} P. A. M. Dirac, Proc. Roy. Soc. {\bf A209}, 291 (1951);
{\bf 212}, 330 (1951); {\bf 223}, 438 (1954).

\bibitem{Clayton3} M. A. Clayton, (private communication).

\bibitem{Weinberg2} S. Weinberg, {\it Gravitation and Cosmology: Principles
and Applications of the General Theory of Relativity}, John Wiley \& Sons,
New York, 1972.

\bibitem{Penrose} R. Penrose, {\it General Relativity, an Einstein
Centennary Survey}, ed. by S. W. Hawking and W. Israel, Cambridge
University Press (Cambridge, 1979).

\bibitem{Netterfield} C. B. Netterfield et al. astro-ph/0104460; C. Pryke
et al. astro-ph/0104490; R. Stompor et al. astro-ph/0105062.

\bibitem{Perlmutter} S. Perlmutter et al. Ap. J. \textbf{483}, 565 (1997),
astro-ph/9608192; A. G. Riess, et al. Astron. J. \textbf{116}, 1009 (1998),
astro-ph/9805201; P. M. Garnavich, et al. Ap. J. \textbf{509}, 74 (1998),
astro-ph/9806396; S. Perlmutter et al. Ap. J. \textbf{517}, 565 (1999),
astro-ph/9812133; A. G. Riess, et al., astro-ph/0001384.

\bibitem{Brandenberger} R. H. Brandenberger, astro-ph/0208103.

\bibitem{Bond} J. L. Sievers et al., astro-ph/0205387.

\bibitem{Brandenberger2} See e.g.: V. F. Mukhanov, H. A. Feldman and R. H.
Brandenberger, Phys. Rept. {\bf 215}, 20 (1992).


\end{thebibliography}
\end{document}